\documentclass[conference]{IEEEtran}
\IEEEoverridecommandlockouts
\usepackage[numbers]{natbib}
\usepackage{graphicx}
\usepackage{amsmath}
\usepackage{amsthm}
\usepackage{amssymb}
\usepackage{graphicx}
\usepackage{url}
\usepackage{soul}
\usepackage{acronym}
\usepackage{proof}
\usepackage{wasysym}
\usepackage{fancyvrb}
\usepackage{todonotes}
\usepackage{textcomp}
\usepackage{amssymb}
\usepackage{relsize}
\usepackage{color}
\usepackage{todonotes}
\usepackage{algorithm}
\usepackage{algorithmic}
\usepackage{float}
\usepackage{setspace}
\usepackage{subfigure}
\usepackage{breakurl}
\usepackage{multirow}
\usepackage{enumitem}
\usepackage{listings}
\usepackage{xcolor}
\usepackage{lipsum}
\usepackage{caption}
\usepackage{url}
\usepackage{hyperref}
\usepackage{flushend}
\setcitestyle{square}

\newcommand{\subparagraph}{}

\hypersetup{
    colorlinks=true,
    linkcolor=blue,
    filecolor=blue,      
    urlcolor=blue,
    citecolor=blue
}
\usepackage[compact]{titlesec}
\titlespacing{\section}{0pt}{*1}{*1}

\newtheorem{definition}{Definition}

\definecolor{darkgreen}{rgb}{0,0.5,0}

\definecolor{grigiomoltochiaro}{gray}{0.97}
\definecolor{verde}{rgb}{0,1,0}
\newcommand{\define}{\triangleq}

\def\BibTeX{{\rm B\kern-.05em{\sc i\kern-.025em b}\kern-.08em
    T\kern-.1667em\lower.7ex\hbox{E}\kern-.125emX}}
\begin{document}

\title{Immutable Autobiography of Smart Cars}


\author{
    \IEEEauthorblockN{Md Sadek Ferdous\IEEEauthorrefmark{1}\IEEEauthorrefmark{2}, Mohammad Jabed Morshed Chowdhury\IEEEauthorrefmark{3}, Kamanashis Biswas\IEEEauthorrefmark{4}, Niaz Chowdhury\IEEEauthorrefmark{5}}
    \IEEEauthorblockA{\IEEEauthorrefmark{1}Shahajalal University of Science and Technology, Sylhet, Bangladesh. sadek-cse@sust.edu}
    \IEEEauthorblockA{\IEEEauthorrefmark{2}Imperial College London, London, UK. s.ferdous@imperial.ac.uk}
    \IEEEauthorblockA{\IEEEauthorrefmark{3}Swinburne University of Technology, Victoria, Australia. mjchowdhury@swin.edu.au}
    \IEEEauthorblockA{\IEEEauthorrefmark{4}Griffith University, Queensland, Australia. k.biswas@griffith.edu.au}
    \IEEEauthorblockA{\IEEEauthorrefmark{5}Open University, Milton Keynes, UK. niaz.chowdhury@open.ac.uk}
}

\maketitle

\begin{abstract}

The popularity of smart cars is increasing around the world as they offer a wide range of services and conveniences. These smart cars are equipped with a variety of sensors generating a large amount of data, many of which are sensitive. Besides, there are multiple parties involved in a lifespan of a smart car, such as manufacturers, car owners, government agencies, and third-party service providers who also produce data about the vehicle. In addition to managing and sharing data amongst these entities in a secure and privacy-friendly way which is a great challenge itself, there exists a trust deficit about some types of data as they remain under the custody of the car owner (e.g. satellite navigation and mileage data) and can easily be manipulated. In this paper, we propose a blockchain supported architecture enabling the owner of a smart car to create an immutable record of every data, called the \textit{autobiography} of a car, generated within its lifespan. We also explain how the trust about this record is guaranteed by the \textit{immutability} characteristic of the blockchain. Furthermore, the paper describes how the proposed architecture enables a secure and privacy-friendly sharing of smart car data between different parties in a secure yet privacy-friendly manner.

\end{abstract}

\begin{IEEEkeywords}
Smart Car, Autobiography, Data Sharing, blockchain.
\end{IEEEkeywords}

\section{Introduction}
\label{sec:intro}
The ever-increasing popularities of smart cars imposes a significant challenge from the viewpoint of the management and the secure and privacy-friendly sharing of the data generated by such vehicles. The reason is manifold. Firstly, multiple parties (e.g., manufacturers, government authorities, service providers) are involved who produce data about the car throughout its lifespan and keep records in their database. Secondly, smart cars are embedded with a wide range of sensors which continuously generate data, but the lack of a proper mechanism to enable the car owners from sharing this information with the third party service providers (e.g. an insurer) creates the `silos' of information for any particular car. Therefore, there is no global or holistic view of the information related to a specific vehicle.

This problem is aggravated when we consider that many of such data is private in nature, e.g. driving habit, driving locations and so on. One of the main reasons why the manufacturers do not allow the sharing of individuals' data is the security and privacy threats such as location tracking or remote hijacking of the car. Therefore, the management and sharing of such data must be accomplished in a secure and privacy-friendly way. We argue that blockchain, a disruptive technology that has found many applications from cryptocurrencies to smart contracts, can be a potential solution to these challenges. 

\textbf{Contributions.} We propose a blockchain-based architecture to allow different stakeholders to join a common platform, to allow owners of smart cars to build a holistic view of immutable records, called \textit{autobiographies}, of their smart cars and to share data with other parties in a secure, transparent and auditable way. We also qualitatively argue the resilience of the architecture against common security attacks using a threat model.

\textbf{Structure.} In Section \ref{sec:background}, we provide a brief description about blockchain and smart car and its underlying issues. Section \ref{sec:mathModel} provides a mathematical model to define what we mean by the term \textit{autobiography} of a smart car. Section \ref{sec:system} presents a threat model, a requirement analysis and the proposed system architecture along with a discussion of probable design issues. Finally, we conclude in Section \ref{sec:conclusion} with hints of future work.

\section{Background}
\label{sec:background}
Two disruptive technologies, \textit{Blockchain} and \textit{Smart Car}, are amongst the most discussed subjects of recent time. These two apparently unrelated topics, however, share a close tie-up that can be capitalized on building the architecture for the next-generation applications to manage and share smart car data in a secure and privacy-friendly manner, what we attempted in this paper. A brief overview of both subjects would help to comprehend the remaining parts of the discussion well, and therefore, is presented afterwards.

\subsection{Blockchain}
\label{sec:back:subsec:blockchain}
Blockchain has gained attention from both industry and academia because of the rapid popularity of the crypto-currencies like Bitcoin \cite{nakamoto2008bitcoin} and Ethereum \cite{dannen2017introducing}. Blockchain is the foundational technology that enables crypto-currencies. It has unique and interesting characteristics, such as immutability, decentralization, data persistence, data provenance, distributed control with accountability and transparency. With all these properties, a blockchain is essentially a distributed, always available, irreversible, tamper resistant, replicated public repository of data where trustless users can agree on an immutable and auditable piece of data without any third party interaction~\cite{alexopoulos2017beyond}. Smart-contract is a mechanism that allows to deploy irreversible and autonomous computational logic on top of blockchain. There are mainly two types of blockchain: a public blockchain is accessible by anyone whereas a private blockchain can set access rules to define who has acesss to the blockchain data.

\subsection{Smart car}
\label{sec:intro:subsec:smartCar}
Smart car is a system, which collects information about its different components and processes these information to provide added-value features to the car owner \cite{enisa2016}. It improves the car owner experience and the car safety. It also enables to communicate with other smart cars and provides different types of telemetries. Understandably, many of the generated data is sensitive in nature. Being a newer domain, it has several issues as illustrated below:

\begin{itemize}
    \item \textbf{Privacy:} Many of the generated data is sensitive in nature.
    \item \textbf{Security:} Apart from the security of the crucial internal components of a smart car, the data generated by smart cars lacks any proper mechanism to guarantee the confidentiality, authenticity and integrity. 
    \item \textbf{Holistic view:} The size of data generated by different sensors of a smart can be huge. Even worse is that there are third parties, e.g. garages, smart infrastructures and so on, who might have data about the smart car. The scattered nature of such data makes it difficult for any owner to have a holistic view during the lifespan of a smart car.
    \item \textbf{Sharing:} The traditional approach makes it difficult to share smart car data in a secure, privacy-friendly and auditable way among different parties.
\end{itemize}

Over the last few years, researchers have published different research works related to security and privacy issues in smart or connected car. For example, in \cite{ishtiaq2010security}, Rouf et. al. have presented the security and privacy vulnerabilities in the wireless network of the smart car. Woo et al have demonstrated a long-range wireless attack using a real vehicle and malicious smartphone application in a connected car environment in \cite{woo2015practical}. Dorri et al. have proposed a blockchain based system to overcome the security and privacy issues in smart car \cite{dorri2017blockchain}. They have demonstrated remote software updates and  dynamic vehicle insurance fees using blockchain. However, they have not provided any mechanism to share car related data with other interested parties or how to build a holistic view of smart car data.

\section{Mathematical Model}
\label{sec:mathModel}

In this section, we present our idea of how we can conceptually represent the holistic view of a smart car data using a novel concept called the \textit{autobiography} of a smart car. To concretize the semantic of the concept, we also build a mathematical model.

As per the ENISA report \cite{enisa2016}, a smart car have the following components where each components have multiple sub-components/sensors .

\begin{itemize}
	\item \textit{Powertrain control:} engine control, transmission control, speed and gear control, power train sensors and so on.
	\item \textit{Chassis control:} Steering control, braking system, ADAS (Advanced Driver Assistance System) systems and so on.
	\item \textit{Body control:} Door locking, air conditioning sensor, light sensors, seatbelt sensor and so on.
	\item \textit{Infotainment control:} Audio-visual unit, navigation, external media unit and so on.
	\item \textit{Communications control:} Gateway ECUs (Engine Control Units), telematics, Communication unit and so on.
	\item \textit{Diagnostic and maintenance systems:} OBD (On-board Diagnostics) ports, external aftermarket dongles and so on.
\end{itemize}

Each of these component/sensors can generate data. Since these components are internal as far as a smart car is concerned, we call them \textit{internal sources}. However, there could be data sources which are \textit{external}, such as a garage providing a yearly fitness certificate, Government transport authority providing a yearly tax certificate as well as registration certificates and even other vehicles (enabling Vehicle to Vehicle communication) or infrastructures (enabling Vehicle to Infrastructure communication).

Now, we gradually build the mathematical model by denoting the set of smart cars with $\mathit{VEC}$. We develop the mathematical model from the perspective of a smart car $\mathit{vec} \in \mathit{VEC}$. The set of other cars is presented using the notation $VEC'$, such that $\mathit{VEC}' \define \{\mathit{VEC} \setminus \mathit{vec} \}$.

We use the notation $S_{\mathit{vec}}$ to denote the set of internal sources of the smart car $\mathit{vec}$. The external sources, as we consider within the scope of this paper, consist of other vehicles (belonging to the set $\mathit{VEC}'$), set of infrastructures (denoted with $\mathit{INF}$) and set of organizations (denoted with $O$). Garages and other regulatory bodies are examples of different organizations within the scope of this paper. Therefore, the set of external sources, denoted with $S_e$ is defined as: $S_e \define \{\mathit{VEC}' \cup \mathit{INF} \cup O\}$.

Then, we can define $S^{\mathit{vec}}$ (the set of all sources) from the perspective of the smart car $vec$ as the union of its internal and external sources. Thus,

$$
S^{\mathit{vec}} \define \{{S_{\mathit{vec}}} \cup {S_e} \}
$$

Based on this foundation, we define the concepts of claim and assertion from the perspective of a smart car:

\begin{definition}\label{claim-def}
\textbf{Claim:} A claim is a statement about a smart car consisting of information, relevant to that smart car, generated either by one of the internal sources or by an external source. The information, thus, can be anything generated by a single sensor. Alternatively, this information can also be generated by another vehicle, by an infrastructure (e.g. traffic light) or by any regulatory or authorized entity. 
\end{definition}

\begin{definition}\label{assertion-def}
\textbf{Assertion:} An assertion is a signed collection of claims. We differentiate between two types of assertions: internal or external. An internal assertion is the collection of claims where each claim is generated by one of the internal sources of the car and signed by the private key of the smart car. On the other hand, an external assertion is a collection of claims produced by one of the external sources and subsequently signed by the respective private key of the source.
\end{definition}

Practically, a claim can be represented using a structured data consisting of name-value pairs where a name presents a property and the value its corresponding information. This concept of structured data can be utilized to formalize the notion of a claim. Let $C^s_{\mathit{vec}}$ denote the set of claims for a car $vec$ generated by an internal source $s \in S_{\mathit{vec}}$. Then, a claim $c \in C^s_{\mathit{vec}}$ can be defined in the following way:

$$
c \define \langle\ (n_1,v_1), (n_2,v_2), (n_3,v_3) ... (n_j,v_j) \, \,\rangle
$$ 

Here, $n$ represents the name of the property, $v$ presents its value and $j \in \mathbb{N}$.

\begin{figure}[!h]
\includegraphics[width=9cm,keepaspectratio]{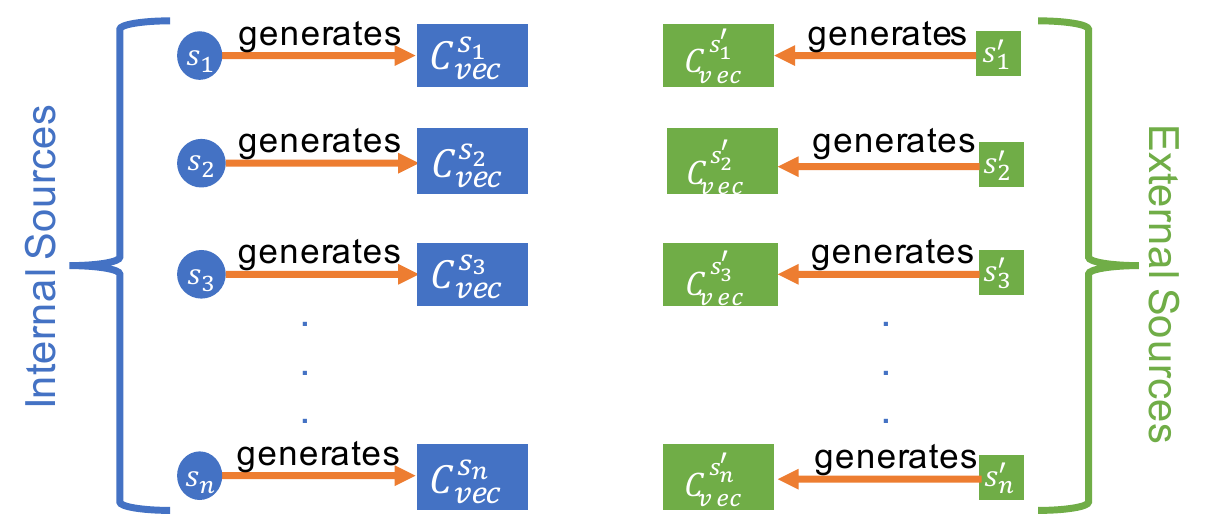}
\centering
\caption{Claims from different sources}
\label{Fig:sources}
\end{figure}

Similarly, we can denote the set of claims for a car $\mathit{vec}$ generated by an external source $s' \in S_e$ using the notation: $C^{s'}_{\mathit{vec}}$ where each single claim consists of name-value pairs as defined before for an external source. A visual illustration of claims is presented in Figure \ref{Fig:sources}.

An internal assertion, denoted as $X_{\mathit{vec}}$ for a car $\mathit{vec}$, can be modelled as a set containing the signed union of all claims generated by the internal sources of a car and defined it in the following way:

$$
X_{\mathit{vec}} \define \bigg\langle\ \bigg\{ {\bigcup\limits_{s \in {S_{\mathit{vec}}}} C^s_{\mathit{vec}}}\bigg\}_{K^{-1}_{\mathit{vec}}}\ \ \bigg\rangle
$$

Here, ${K^{-1}_{\mathit{vec}}}$ represents the private key of the vehicle $\mathit{vec}$.

Finally, we can define an assertion provided by an external source ($s' \in S_e$) regarding a vehicle $\mathit{vec}$ in the following way:

$$
X^{s'}_{\mathit{vec}} \define \bigg\langle\ \bigg\{ C^{s'}_{\mathit{vec}}\bigg\}_{K^{-1}_{s'}}\ \ \bigg\rangle
$$

Next, we provide a definition of the autobiography of a smart-car:

\begin{definition}\label{biography-def}
\textbf{Autobiography:} The autobiography of a smart is defined as the collections of internal and external assertions generated throughout its lifespan.
\end{definition}

Denoted with $\mathit{BIO}_{\mathit{vec}}$ to represent the autobiography of a smart car $vec$, we define it in the following way:

$$
\mathit{BIO}_{\mathit{vec}} \define \bigg\langle\ X_{\mathit{vec}} \bigcup  \bigg\{{\bigcup\limits_{s' \in {S_e}} X^{s'}_{\mathit{vec}}}\bigg\}\ \ \bigg\rangle
$$

\begin{figure}[!h]
\includegraphics[width=7cm,keepaspectratio]{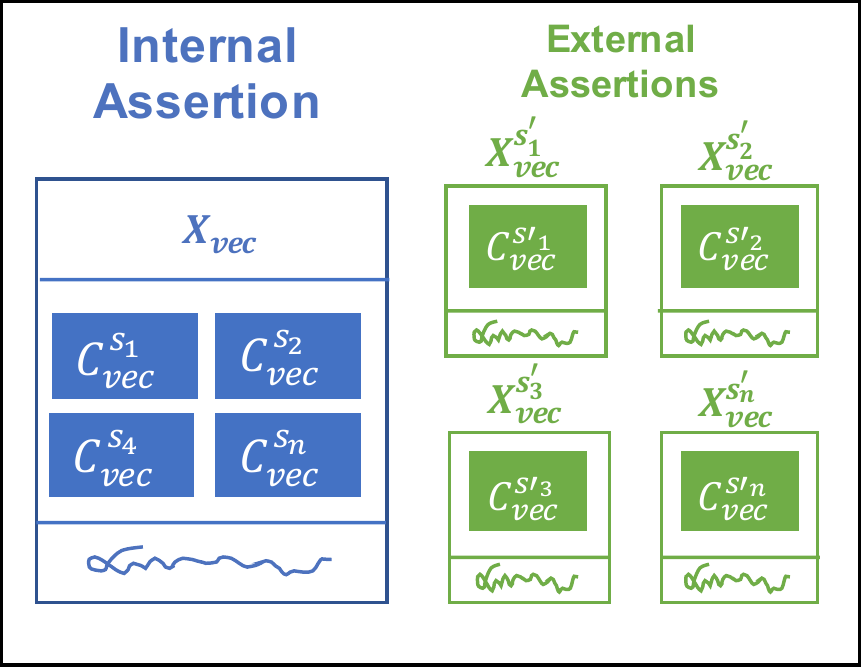}
\centering
\caption{Modelled autobiography of a smart car}
\label{Fig:auto}
\end{figure}

This definition is visually presented in Figure \ref{Fig:auto}

According of this definition of the autobiography, it consists of every piece of information generated by every single internal and external source over the lifespan of a car. Understandably, this can be a large collection of information and might be unsuitable or even not required for exchange and sharing with parties. For such scenarios, we introduce a novel terminology called \textit{Snapshot}, defined in the following way:
\begin{definition}\label{snapshot-def}
\textbf{Snapshot:} The snapshot of a smart car (denoted with $\mathit{SNAP}_{\mathit{vec}}$ for a smart car $\mathit{vec}$) is the collections of internal and external assertions generated with only the required information from the respective sources.
\end{definition}

Essentially, it is created with the subset of information generated and signed by the respective source. Intuitively, it can be modelled in the following way:

$$
\mathit{SNAP}_{\mathit{vec}} \define \bigg\langle\ \mathcal{X}_{\mathit{vec}} \bigcup  \bigg\{\bigcup\limits_{s' \in {S_e}} \mathcal{X}^{s'}_{\mathit{vec}}\bigg\}\ \ \bigg\rangle
$$

Here, $\mathcal{X}_{\mathit{vec}} \subseteq X_{\mathit{vec}}$ and $\mathcal{X}^{s'}_{\mathit{vec}} \subseteq {X}^{s'}_{\mathit{vec}}$.

\section{Proposed System}
\label{sec:system}
In this section, we present the proposed system with the sole purpose to tackle the identified issues. We start with a threat model (Section~\ref{sec:system:subsec:threat}) and requirement analysis to mitigate such threats (Section~\ref{sec:system:subsec:req}), then propose an architecture (Section~\ref{sec:system:subsec:archi}) and finally analyze the architecture (Section~\ref{sec:system:subsec:choice}).
\subsection{Threat Model}
\label{sec:system:subsec:threat}


Threat  modelling is one of the mostly used methods to identify, communicate, and understand threats and mitigation mechanisms within the context of protecting (IT) assets, smart car in the context of this paper. However, it is highly dependant on the capabilities of the attacker that is assumed for a particular system. Traditionally, the capabilities of an attackers are presented using an adversary model, One of the most well known adversary models is the Honest-But-Curious model~\cite{paverd2014modelling} which assumes that an attacker participates in the protocol of the system and behaves correctly. She can intercept, send (or receive) any message to (or from) the protocol to which it participates. However, she cannot modify any message or launch other types of attacks. We believe this is quite restrictive in the sense that modern attackers can exhibit additional capabilities.

We also differentiate between two types of attackers namely \textit{internal attacker} who is the owner of the car motivated to gain illicit advantages by fabricating sensor generated data (e.g. odometer) and \textit{external attacker} who can be either an individual other than the owner or an organization whose main motivation is to act maliciously by getting hold of data from the smart car or disrupting its system or other external systems which might deal with the autobiography of the smart car. With this conjecture in mind, we also assume that an attacker can i) intercept as well as modify any network packets transmitted over an insecure communication channel, ii) disrupt the internal system of the smart car or other external systems which deal with the autobiography of the smart car, iii) gain unauthorized access only if she has access to the required credential, e.g. username/password pair or secret key, and iv) try to alter data generated by different sensors stored within a smart car which ultimately will alter the autobiography of a smart car.

Considering the above capabilities, we have chosen a well established threat model called STRIDE \cite{shostack2014} developed by Microsoft. The STRIDE model is briefly presented below.
\begin{itemize}
    \item \textbf{T1-Spoofing Identity: } The act of spoofing refers to an adversary using the identity of an authorized user (e.g. owner) to illegally access or modify resources on a system where they would normally do not have any access.
    
    \item \textbf{T2-Tampering with Data: } As many of the user data remains under the control of the users, they can change the value of the data (e.g., odometer reading). Therefore, the integrity of the data can be lost.
    
   \item \textbf{T3-Repudiation: } This involves a user or attacker who leverages the inability of a system to track invalid and illegal actions and uses them to gain some advantages in the system. 

    \item \textbf{T4-Information Disclosure:}  Private or sensitive data stored in a smart car may be leaked when the car is sold to another user.

    
    \item \textbf{T5-Denial of Service: } The system that is used to access the smart car data or to share the autobiography can be the target of a denial of service attack.
    
     \item \textbf{T6-Elevation of privilege: } Malicious software with potential exploitable vulnerabilities may be the first step to gain access to external systems (e.g. systems utilized by garages or other TTPs), thus potentially gaining a privileged access on a large set of vehicles. 
\end{itemize}

Among all these threats, we exclude the elevation of privilege from our consideration for this paper. This is because such a threat is more relevant for enterprise systems. Besides, with the involvement of very sensitive private data handled by smart cars, we must also consider the privacy threats related to the lack of control, an addition to the standard STRIDE model. 

\begin{itemize}
    \item \textbf{T7-Lack of control:} The owner has no or little control regarding how data of the smart car is shared with other entities.
\end{itemize}



\subsection{Requirement Analysis}
\label{sec:system:subsec:req}
In this section, we present a set of security and privacy requirements which can be utilized to mitigate the identified threats. The security requirements are:
\begin{itemize}
    \item \textbf{S1:} The data generated in a smart car must be stored in a secure way with a guarantee of its confidentiality and integrity.
    \item \textbf{S2:} Data relevant to the smart car must be transmitted over a secure channel.
    \item \textbf{S3:} There must be a secure way to access the data and create an autobiography with strong guarantee of its integrity. This mitigates \textit{T1}.
    \item \textbf{S4:} The sharing of the car data must be carried out in a secure, transparent and accountable fashion. \textit{S4} in combination with \textit{S1} can tackle \textit{T4}.
    \item \textbf{S5:} The authenticity and non-repudiation of data must be guaranteed. \textit{S5} in combination with \textit{S1}, \textit{S2} and \textit{S3} can mitigate \textit{T2} and \textit{T3}.
    \item \textbf{S6:} A distributed system should be leveraged in order to minimize the impact of any DoS attack on the relevant system. This tackles \textit{T5}.
\end{itemize}

Next, the privacy requirements are presented:

\begin{itemize}
    \item \textbf{P1:} The sharing platform must satisfy the selective disclosure property, allowing the owner to build an autobiography with the full control over the data.
    \item \textbf{P2:} The autobiography must be shared with the explicit consent of the owner. \textit{P1} and \textit{P2} combinedly can mitigate \textit{T7}.
\end{itemize}

\subsection{Architecture}
\label{sec:system:subsec:archi}

\begin{figure*}[!h]
\includegraphics[width=12cm,keepaspectratio]{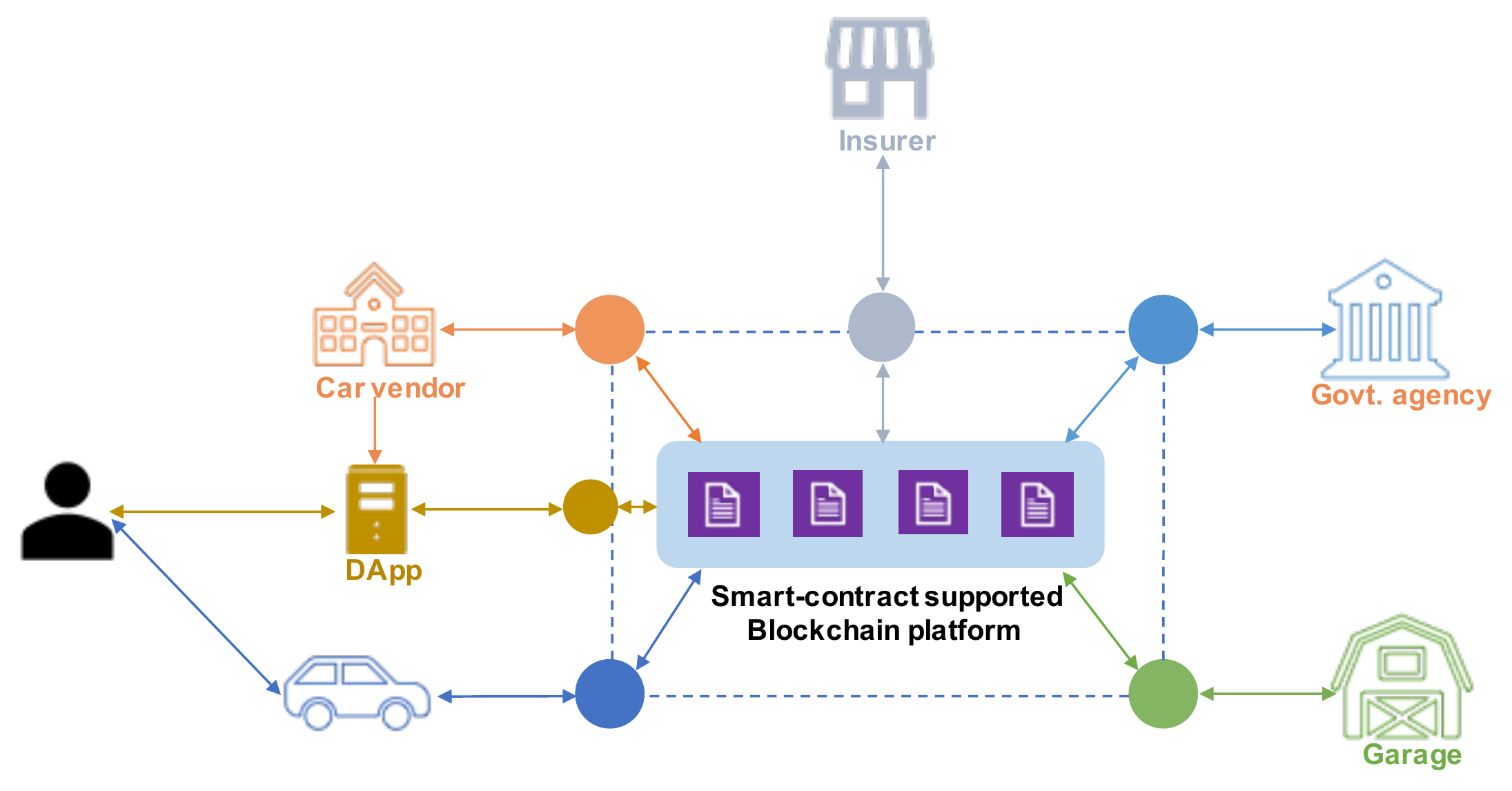}
\centering
\caption{Architecture of the proposed platform}
\label{Fig:archi}
\end{figure*}

The architecture of the proposed platform is presented in Figure \ref{Fig:archi}. A brief overview of different aspects of the architecture is provided next.

The architecture is centered around a smart-contract supported blockchain platform. Different nodes connected to the blockchain network are the smart car vendors, smart cars, different government regulatory agencies, car insurers and other third party service providers such as garages. The car vendor during the production of the car (denoted as $\mathit{vec} \in \mathit{VEC}$) will generate a key-pair (public key $K_{\mathit{vec}}$ and private key $K^{-1}_{\mathit{vec}}$) utilizing the blockchain platform. The private key is stored in a tamper-proof hardware of the smart car while the public key is included as a digital certificate signed by the vendor.

A smart car generates different claims in its lifespan. However, for simplicity, we make the assumptions that i) a smart car is always connected to the network, ii) a claim is generated in a specified duration where and is stored in the internal storage of the smart car and its hash is stored in the blockchain via a transaction signed by the private key of the car, and iii) an internal assertion is created when a trip is completed where the assertion is stored in the internal storage of the car and its hash is stored in the blockchain via a transaction which is signed by the private key of the car.

The owner of a smart car can always connect to the smart car to download different claims and assertions either via a web, mobile or a desktop app and store them securely in encrypted formats in an external storage of the owner. We assume that the car vendor provides such an application along with a decentralized application (the so called \textit{DApp}) to interface with the blockchain platform. Whenever such data is downloaded from the smart car, the integrity of each data is checked with its corresponding hash from the blockchain. If the hashes do not match, it implies that the data stored in the smart car has been altered and hence, an error message is displayed to the user and a marker is set to signify this.

Intuitively, a smart contract will be utilized for each user by which different information (mostly the hashes) regarding the respective smart car will be stored or retrieved. In addition, other nodes will also utilize their own smart contract to facilitate different functionalities. One of the core functionalities is the sharing of assertions, e.g. a tax certificate and a fitness certificate, between different entities within the platform. We envision the following way for the transfer of external assertions to the owner of a smart car.

We assume a \textit{Repository} smart contract that holds the identity of the car, its corresponding public key and the address of the smart contract for all relevant entities within the network. In such, this smart contract will be the one-stop entry to retrieve the identity-public key pair or the identity-smart contract address pair for a particular entity in the network. This Repository contract can be maintained by the vendor itself or by any third party.

When an external third party would like to share an external assertion it can retrieve required information, public key and address of the respective smart contract, by just looking up in the Repository smart contract. Then, the third party can just create an assertion, optionally encrypt it with the public key of the car and then send over the respective smart contract of the car via a signed transaction. Once the car smart contract receives the transaction, it notifies the owner and upon being approved, the assertion is then stored in the smart contract, which later can be retrieved using any preferred device.

The owner can use an app to combine different internal and external assertions ($X_{\mathit{vec}}$ and $X^{S'}_{\mathit{vec}}$) into an autobiography ($\mathit{BIO}_{vec}$) of the smart car with its hash stored in the blockchain. Eventually, when the autobiography is re-updated, the hash is also updated.

The size of such an autobiography will be a significant issue when data from different sensors is combined. Additionally, having the whole spectrum of data into the autobiography will raise different privacy questions. That is why it might not be ideal to share it with an external entity (e.g. an insurer). Towards that aim, a user can choose the information that is to be shared to create a snapshot of the autobiography ($\mathit{SNAP}_{vec}$). The owner then looks into the Repository smart contract to retrieve the public key and smart contract address of the external entity and sends the snapshopt to the smart contract. The snapshopt can be encrypted for additional security using the public key of the external entity. Once the smart contract of the external entity receives the snapshot it goes through the same verification process as described before. 

When the ownership of a smart car changes, the architecture proposes a special type of ownership transactions, initiated by the new owner and verified by a smart contract of a Government agency, which would do the following steps:
\begin{itemize}
    \item Create an assertion from an aggregated version of all claims from the car as generated while in the possession of the previous owner in a privacy friendly way. This is to ensure that the new owner cannot get hold of sensitive private data (e.g. driving locations) of the old owner while preserving the essence of the usage of the car. One example of this is that the new assertion contains the total mileage of the smart car under the old owner instead of fine-grained details of where, when and for how long the car was driven previously.
    \item Import all data (mainly assertions) from the corresponding smart contract under the old owner and create a new smart contract under the new owner.
    \item Destroy the old smart contract in such a way that there is no way to retrieve data from it.
\end{itemize}
These steps will ensure that the important information regarding the smart car does not get lost with the change of ownership while preserving the privacy of the old owner.
\subsection{Analysis \& Design Choice}
\label{sec:system:subsec:choice}
In this section we analyze the ways the proposed architecture can satisfy different requirements. Data generated by different sensors of the smart car is stored, as per the proposal, either in a temper proof hardware while in the car storage or in encrypted format when downloaded in an external storage or stored in the different smart contracts in the form of assertions, autobiography or snapshot. In addition their hashes are stored in the blockchain, guaranteeing its integrity. All these satisfy the \textit{S1} requirement. 

The protocol for the architecture can be deployed in such a way that data always is transmitted over an encrypted channel, thereby satisfying \textit{S2}. To ensure \textit{S3}, the desktop, web or mobile app should be equipped with a strong authentication mechanism so that only the owner, or an authorized delegated entity, can access as well as generate the autobiography or the snapshot involving the car data. The proposal requires that the hashes of the assertions as well as autobiography and snapshot are stored in the blockchain. This creates an immutable evidence for the corresponding element which satisfies \textit{S3}.

The sharing of assertions and snapshots is always carried out within the blockchain via signed transactions. This creates a transparent and auditable trails while ensuring the security and thus satisfies \textit{S4}. The signed assertions and transactions provide a strong guarantee of authenticity and non-repudiation as it is assumed only the authorized owner can utilize the corresponding private keys. This satisfies \textit{S5}. An added advantage of a blockchain-supported platform is its decentralization capabilities. This minimizes the impact of any DoS attack and hence satisfies \textit{S6}.


The proposed system advocates that each internal assertion and the snapshot is created with the involvement and explicit consent of the owner with the ultimate authority to decide what information is to be included in such elements. This satisfies \textit{P1} and \textit{P2}. To realize such a proposal, there are several design choices that need to be finalized. Firstly, the type of a blockchain platform that needs to be utilized for the proposal. We envision a private blockchain platform such as Hyperledger Fabric \cite{fabric2018}. This is because a private blockchain offers a better support of privacy, scalability and throughput. Secondly, the next issue is the responsibility to maintain such a blockchain. We believe a consortium of car vendors, Government bodies and third parties, as outlined in our paper, can be an good choice. This will ensure a shared responsibility between the inter-connected parties and greater benefits for everyone involved within this ecosystem.

\section{Conclusion}
\label{sec:conclusion}
Privacy and security have been becoming more critical as modern cars are getting smarter. The smart cars themselves and participating organizations generate data about the vehicle that remain under the custody of different parties including the car owner. Building trust about the data, securing storage and sharing data in a privacy-preserving manner are amongst the main future challenges. In this paper, we have proposed a system architecture that can address these three challenges using blockchain technologies and a novel concept called the ``autobiography of a smart car''. The idea has been mathematically formalized to concretize its semantic. A threat model is used to derive the system requirements for our proposed system, and finally, we have presented and discussed our system architecture and how it meets all those identified requirements. Overall, if our system is deployed, it will provide a holistic view of data about any particular car, and enable car owners to share their data with others in a privacy-friendly way. We believe that it will pave down the way for future research in this domain.


  
\end{document}